# Broadband super-Planckian thermal emission from hyperbolic metamaterials

Yu Guo, Cristian L Cortes, Sean Molesky and Zubin Jacob*
Department of Electrical and Computer Engineering
University of Alberta, Edmonton, AB T6G 2V4, Canada
zjacob@ualberta.ca

**Abstract**: We develop the fluctuational electrodynamics of metamaterials with hyperbolic dispersion and show the existence of broadband thermal emission beyond the black body limit in the near field. This arises due to the thermal excitation of unique bulk metamaterial modes, which do not occur in conventional media. We consider a practical realization of the hyperbolic metamaterial and estimate that the effect will be observable using the characteristic dispersion (topological transitions) of the metamaterial states. Our work paves the way for engineering the near-field thermal emission using metamaterials.

Engineering the black body thermal emission using artificial media promises to impact a variety of applications involving energy harvesting[1], thermal management[2] and coherent thermal sources[3]. The usual upper limit to the black-body emission is not fundamental and arises since energy is carried to the far-field only by propagating waves emanating from the heated source. If one allows for energy transport in the near-field using evanescent waves, this limit can be overcome. Thus thermal emission beyond the black body limit is expected due to surface electromagnetic excitations[4] or at the edge of the bandgap in photonic crystals[5] where there is a large enhancement in the photonic density of states. Advances in near field scanning and probing techniques[6–8] have led to conclusive demonstrations of these effects.

One limitation of the above mentioned approaches using photonic crystals or surface electromagnetic excitations is that the energy transfer beyond the black body limit (super-planckian[9,10] thermal emission) only occurs in a narrow bandwidth. In this paper, we show that artificial media (metamaterials) with engineered dielectric properties can overcome the limitation of super-planckian thermal emission at a single resonant frequency. Our work rests on the recently discovered singularity in the bulk density of states of metamaterials with hyperbolic dispersion[11–13]. The unique property which sets these hyperbolic metamaterials (HMMs) apart from conventional approaches of engineering the photonic density of states (PDOS) is the broad bandwidth in which the PDOS is enhanced.

The dielectric tensor of the HMM is given by $\ddot{\varepsilon} = diag[\varepsilon_{xx}, \varepsilon_{yy}, \varepsilon_{zz}]$ with $\varepsilon_{xx} = \varepsilon_{yy}$ and $\varepsilon_{xx} \cdot \varepsilon_{zz} < 0$. The extraordinary waves in this extremely anisotropic uniaxial medium follow a hyperboloidal isofrequency surface $(k_x^2 + k_y^2)/\varepsilon_{zz} + k_z^2/\varepsilon_{xx} = k_0^2$ ($\vec{k} = [k_x, k_y, k_z]$ and $k_0$ is the free space wavector) as opposed to spherical in a conventional medium (Fig. 1). It follows immediately that waves with very large wavevectors are allowed to propagate in this HMM which would have have been evanescent in any conventional medium (Fig. 1(a). red arrows). The PDOS enhancement occurs due to the contribution of these high-$k$ modes. The use of these states for quantum nanophotonics was proposed[11] and later experimentally verified by multiple groups[14–16].

Here, we develop the fluctuational electrodynamics of HMMs to analyze the role of the large enhancement in the density of states on the thermal emission. Our work paves the way for engineering the thermal emission using metamaterials for applications such as near-field thermophotovoltaics[17] and near-field thermal sources. We show that the characteristic dispersion of the high-$k$ states (topological transitions in the isofrequency surface[18]) manifests itself in the thermal emission beyond the black body limit. We also show that the proposed effect can be experimentally verified in a practical implementation of the HMM.

One approach to achieving the extreme anisotropy characteristic of HMMs is by using a metal-dielectric super-lattice structure where the thickness of each layer is far below the operating wavelength. We consider here a phonon-polaritonic metal (silicon carbide SiC) where the coupling between the optical phonons and light causes the real part of the permittivity to become negative in the Reststrahlen band (real($\varepsilon_m$) < 0 between $\omega_{TO} = 149.5 \times 10^{12}$ Hz and $\omega_{LO} = 182.7 \times 10^{12}$ Hz, the transverse and longitudinal optical phonon resonance frequencies). The permittivity of SiC is given by[4] $\varepsilon_m = \varepsilon_\infty (\omega_{LO}^2 - \omega^2 - i\gamma\omega)/(\omega_{TO}^2 - \omega^2 - i\gamma\omega)$ where $\omega$ is the frequency of operation, $\varepsilon_\infty = 6.7$ and $\gamma = 0.9 \times 10^{12}$ Hz. The key advantage of these phonon-polaritonic HMMs, along with the low losses, is that the high-$k$ modes can be excited thermally at moderate temperatures ranging from 400-500 K when the black body distribution lying within the Reststrahlen band takes on experimentally detectable values. We choose silicon dioxide

(SiO$_2$) as the dielectric which is compatible with SiC and can be epitaxially grown in a chemical vapor deposition process[19] ($\varepsilon_d = 3.9$).

We now introduce the possible optical phases of the SiC/SiO$_2$ based multilayer hyperbolic metamaterial. Effective medium theory predicts the components of the metamaterial along the direction parallel and perpendicular to the interfaces to take the form $\varepsilon_{xx} = \varepsilon_{yy} = \varepsilon_m f + \varepsilon_d(1-f)$ and $\varepsilon_{zz} = \varepsilon_m \varepsilon_d / (\varepsilon_m(1-f) + \varepsilon_d f)$ respectively, where $f$ is the fill fraction of the metal[16]. When the fill fraction is changed, the composite metal-dielectric multilayer slab behaves as a type I hyperbolic metamaterial with only one negative component in the dielectric tensor ($\varepsilon_{xx} = \varepsilon_{yy} > 0, \varepsilon_{zz} < 0$), type II hyperbolic metamaterial with two negative components ($\varepsilon_{xx} = \varepsilon_{yy} < 0, \varepsilon_{zz} > 0$), effective anisotropic dielectric ($\varepsilon_{xx} = \varepsilon_{yy} > 0, \varepsilon_{zz} > 0$) or effective anisotropic metal ($\varepsilon_{xx} = \varepsilon_{yy} < 0, \varepsilon_{zz} < 0$)[16]. The optical phases are defined with respect to the extraordinary waves in the metamaterial. The various optical phases as a function of wavelength and fill-fraction for the SiC/SiO$_2$ multilayer phonon-polaritonic HMM have been shown in Figure 1 along with their isofrequency surfaces. Note that both HMMs support propagating high-$k$ modes which do not occur in conventional media. Furthermore, as the wavelength is changed, there are topological transitions in the isofrequency surfaces (closed to open or vice versa) marked by the sudden appearance of the high-$k$ states[18].

We now explain the effect of these high-$k$ modes and optical phases on thermal emission from HMMs. To calculate the thermal emission in the far-field, Kirchoff's law suffices, which states that the spectral directional absorptivity is equal to the spectral directional emissivity[20]. Here, we focus exclusively on the near field properties which cannot be accounted for by Kirchoff's laws. Note that the high-$k$ waves which are thermally excited in the HMM are trapped inside and will be evanescent in vacuum (not reach the far field).

Following the seminal work of Rytov[21], which considers electromagnetic fields generated by equilibrium thermal fluctuations, we use the fluctuation-dissipation theorem and the green's tensor of the hyperbolic metamaterial (see supplementary information) to calculate the emitted energy density in the near field of an HMM in equilibrium at temperature T

$$u(z,\omega,T) = \frac{U_{BB}(\omega,T)}{2}\left\{\int_0^{k_0} \frac{k_\rho dk_\rho}{k_0|k_{1z}|} \frac{(1-|r^s|^2)+(1-|r^p|^2)}{2} \right.$$
$$\left. + \int_{k_0}^\infty \frac{k_\rho^3 dk_\rho}{k_0^3|k_{1z}|} e^{-2\operatorname{Im}(k_{1z})z}(\operatorname{Im}(r^s)+\operatorname{Im}(r^p))\right\} \quad (1)$$

where $u(z,\omega,T)$ is the energy density at a distance z and frequency $\omega$ while T denotes the temperature. $k_\rho = \sqrt{k_x^2 + k_y^2}$, $k_0 = \omega/c$, $k_{1z} = \sqrt{k_0^2 - k_\rho^2}$ (Im $k_{1z} > 0$), $r^s$ and $r^p$ are the Fresnel reflection coefficients from the metamaterial half space for (s) and (p) polarized light respectively and "Im" denotes the imaginary part. $U_{BB}(\omega,T)$ is the black-body emission spectrum at temperature T. For convenience, we choose a metamaterial half space but we have verified that our approach and also the results can be generalized to a multilayer slab or effective medium slab. These results will be presented in a detailed publication elsewhere.

Notice that the second term in the above equation is the contribution to the near field energy density due to evanescent waves ($k_\rho > k_0 = \omega/c$) and is the reason for the super-planckian emission. In the near field when $d \ll \lambda$, the above expression simplifies to

$$u(z,\omega,T)^{z\ll\lambda} \approx U_{BB}(\omega,T)\frac{\operatorname{Im}(r^p)}{8(k_0 z)^3} \quad (2)$$

Near the HMM ($\varepsilon_\| = \varepsilon_{xx} + i\varepsilon''$, $\varepsilon_\perp = \varepsilon_{zz} + i\varepsilon''$ and $\varepsilon_{xy} \cdot \varepsilon_{zz} < 0$), the reflection coefficient of high-k waves takes the form $r^p \approx (\varepsilon_\| - \sqrt{\varepsilon_\|/\varepsilon_\perp})/(\varepsilon_\| + \sqrt{\varepsilon_\|/\varepsilon_\perp})$ and thus finally we have the near-field energy (for low losses) to be

$$u(z,\omega,T)^{z\ll\lambda} \approx \frac{U_{BB}(\omega,T)}{8}\left[\frac{2\sqrt{|\varepsilon_{xx}\varepsilon_{zz}|}}{(k_0 z)^3(1+|\varepsilon_{xx}\varepsilon_{zz}|)} - \varepsilon''\frac{2(\varepsilon_{xx}+\varepsilon_{zz})}{(k_0 z)^3(1+|\varepsilon_{xx}\varepsilon_{zz}|)^2}\right] \quad (3)$$

The second term which depends on the imaginary part of the dielectric constant is due to the well known non-radiative local density of states enhancement ("lossy surface waves") in the near field of any absorptive body but the first term is solely due to the bulk hyperbolic metamaterial high-k modes[16,18]. Note that this term dominates the thermal emission in the near field and does not exist for any conventional medium[11]. Furthermore as we will show using a complete calculation, the comparatively lower losses in phonon-polaritonic metals as opposed to plasmonic metals makes the effect exceptionally strong as compared to lossy surface waves.

We now calculate this near field energy density (Eq. 2) for the entire optical phase space of the SiC/SiO$_2$ hyperbolic metamaterial. A remarkable agreement is seen between the optical phases in Fig. 1 and the near field thermal emission in Fig. 2. A number of interesting features are observed in the various phases. The thermal emission exceeds the black body limit in three regions a) type I HMM b) type II HMM and c) dark red line in the effective metal region which is due to a surface-phonon-polariton resonance. We note that c) occurs only in a narrowband range in the effective anisotropic metal region[22] and does not occur for fill fraction less than 0.5 (the resonance condition is $\varepsilon_\| \cdot \varepsilon_\perp = 1$ which cannot be fulfilled unless we have an effective metal). We therefore focus our discussions on metamaterials with $f \leq 0.5$ to emphasize the role of the high-$k$ modes.

To understand whether the analytical high-$k$ approximation is realized in practice, we perform a full calculation for metamaterials with fill fraction 0.5 and 0.25. The results of the full calculation are in excellent agreement with our high-$k$ approximation at distances of z=100 nm which can easily be probed in experiment (Fig. 3). Notice that the two regions of broadband super-planckian thermal emission (normalized to the black body emission into the upper half space) are due to type I and type II hyperbolic metamaterials (Fig 3(a)). The metal-dielectric composite behaves like an effective dielectric in the region in between these two phases and hence the thermal emission is low (no high-$k$ modes). When the fill fraction is 0.5, these two frequency ranges are not separated by bands of effective dielectric or metal and hence super-planckian thermal emission is observed in the entire Reststrahlen band of silicon carbide (Fig. 3(b)). These effects can be classified as the consequence of the metamaterial topological transitions[18] on thermal emission.

The two small peaks in Fig. 3(b) are related to the transverse optical phonon resonance of silicon carbide ($\omega_{TO}$) and the resonant meeting point of all the optical phases ($\omega_s = 171.2 \times 10^{12} Hz$) where the EMT parameters reach a singular limit ($\varepsilon_{xx} = \varepsilon_{yy} \to 0, |\varepsilon_{zz}| \to \infty$) (Fig 3(b) inset).

A natural question arises whether the optical phases predicted by effective medium theory can be achieved in practice. We first note that the wavelength of operation of the metamaterials lies between 10-12 $\mu m$ and hence it is relatively easy to fabricate extremely subwavelength layers (a= 50-100 nm), where a is the unit cell size. We therefore expect the effective medium theory to be valid. To verify this expectation, we calculate the local density of electromagnetic states[23,24] (LDOS) in the near-field of the hyperbolic metamaterials using a practical multilayer realization[16]. The LDOS ($\rho^E(z,\omega)$) is related to processes such as spontaneous and thermal emission. Our simulations take into account the role of losses, dispersion, finite unit cell size, finite sample size and also effects due to the substrate.

We define and calculate the wavevector-resolved local density of states (WLDOS) which carefully elucidates the contribution of the high-$k$ waves to the near field local density of states and is related to the thermal emission[16]. The WLDOS ($\rho^E(z,\omega,k_\rho)$) can be obtained from the Green's tensor[25] as follows $\rho^E(z,\omega) = (2\omega/(\pi c^2))\operatorname{Im}(Tr(\vec{\vec{G}}^E(z,z;\omega))) = \int_{k_\rho} \rho^E(z,\omega,k_\rho)dk_\rho$ (see supplementary information). In figure 4 and 5, we show a false color plot of the available electromagnetic modes in the near field of the metamaterial (fixed distance z=100 nm) as a function of frequency and the lateral wavevector for two different fill fractions ($f = 0.25$ and $f = 0.5$). The bright bands for waves with $k_\rho/k_0 > 1$ correspond to a high density of states due to the coupled phonon-polaritonic bloch waveguide modes which are present in the multilayer realization, in excellent agreement with high-$k$ modes predicted by EMT. Note that waves with such large wavevectors are evanescent in conventional media. Thus they do not carry power and do not contribute to the LDOS.

In figure 4, the fill fraction of the metal is chosen to be 0.25, so that a type I response is achieved from $1.78\times10^{14}$ Hz to $1.82\times10^{14}$ Hz and a type II response is achieved in between $1.49\times10^{14}$ Hz and $1.62\times10^{14}$ Hz. For the multilayer structure, the thickness of the SiC layers are $d_m = 50$ nm and the thickness of the SiO$_2$ layers are $d_d = 150$ nm. The total number of layers are N=10 achieving a net thickness of D=1 $\mu m$. On comparing the effective medium predictions with the practical structure we see an excellent agreement of the optical phases as well as presence of the high-$k$ modes. The multilayer structure as well as the EMT structure behave as an effective dielectric in the dark band where there are no high-*k* modes. Thus the topological transitions predicted in the optical phase diagram are achieved in the practical structure. In figure 5, the fill fraction of the metal is chosen to be 0.5 with $d_m = d_d = 100$ nm, N=10, D=1 $\mu m$ so that a type I response is achieved in $1.71 \times 10^{14}$ Hz to $1.82 \times 10^{14}$ Hz and a type II response is achieved in between $1.49 \times 10^{14}$ Hz and $1.71 \times 10^{14}$ Hz. The LDOS is enhanced over a broad bandwidth as expected from EMT.

The number of high-*k* modes diverges in the EMT limit but we note that there is an upper cut-off for the maximum possible lateral wavevector in the multilayer realization[11,14]. Waves with wavevectors beyond this upper cut-off ($k^{eff} \sim 1/a$, where a is the unit cell size) do not perceive the metamaterial in the effective medium limit and start Bragg scattering. Inspite of this, modes with wavevectors as large as $k_\rho^{max} \approx 50 k_0$ are excited in the practical structure leading to a large local density of states enhancement in the near field (see Fig. 4 and 5). We emphasize that the physical origin of these high-*k* metamaterial bulk waves are the coupled short range surface phonon polaritons at the silicon carbide and silicon dioxide interfaces.

Having ascertained the presence of bulk high-*k* propagating waves in a practical phonon-polaritonic realization of the hyperbolic metamaterial, we now discuss the difference of our approach as compared to surface electromagnetic excitations[4] or photonic crystal structures[5]. The presence of a surface phonon polariton resonance at the interface of air and silicon carbide leads to a large peak in the thermal emission beyond the black body limit. This effect occurs at a frequency specified by the material resonance ($\omega_{sphp} = 178.7 \times 10^{12} Hz$) and hence is narrowband and not tunable. The super-planckian emission presented in this paper originates from bulk metamaterial modes which occur in a broadband frequency range and depends on the fill fraction (thickness) of the metal layers.

Photonic crystal structures can also possess a large photonic density of states at the edge of the bandgap leading to super-planckian thermal emission in a narrow range. The layer thicknesses considered in our paper are far below the size of the wavelength and we function away from the photonic crystal regime to achieve a broadband enhancement in the density of states.

Another effect related to surface excitations are known as "lossy surface waves", which is an enhancement in the non-radiative local density of states in the near field of any absorptive body (dielectric or metal)[4,21,26]. They are known to be the cause of quenching of emitters (eg: dye molecules) near metal surfaces. The effect mentioned in the paper will be easily discernible in experiment from the near-field thermal excitations of such lossy surface modes since they do not show any of the dispersion effects mentioned in our paper. Furthermore, on comparison of the near field thermal emission from silicon carbide (which is a lossy metal or dielectric in the considered frequency range) and the practical hyperbolic metamaterial, we see two orders of magnitude enhancement at a distance of z=100 nm in a broad bandwidth (except at the sphp resonance condition for SiC $\omega_{sphp} = 178.7 \times 10^{12} Hz$). This is explained by the comparatively moderate losses of silicon carbide in the Reststrahlen band.

We emphasize that in all the above cases including hyperbolic metamaterials, the presence of an interface is enough to guarantee that the far-field emissivity is limited to unity[9,10]. To capture the energy in the near field from evanescent waves, a tunneling[8] or scanning[6,7,27] experiment is necessary. Note that the extreme near field limit ($z \ll \lambda$) easily obtainable at mid-infrared wavelengths ($\lambda \approx 10 \mu m$) implies our scaling law will be verifiable experimentally. Furthermore, the deep subwavelength layers achievable without difficult nanofabrication (a = 100 nm $\ll \lambda$) means the effective medium limit will also be attained (as shown by our WLDOS simulations). Experimental measurement of the characteristic dispersion (topological transitions[18]) of the thermal emission due to the metamaterial states for samples with different fill fractions can conclusively support our theoretical predictions.

In conclusion, we have established near field thermal emission exceeding the black body limit from hyperbolic metamaterials in a broad bandwidth. They arise due to bulk high-*k* propagating waves in the metamaterial which cause a divergence of the density of states (effective medium limit)[11]. The predictions of effective medium theory are in excellent agreement with a practical multilayer phonon-polaritonic realization of the metamaterial. Our prediction is verifiable experimentally and we expect our work to pave the way for near-field thermal engineering using hyperbolic metamaterials.

Z. Jacob acknowledges E.E. Narimanov for fruitful discussions. This work was partially supported by the National Science and Engineering Research Council of Canada, Canadian School of Energy and Environment and Alberta Nanobridge.

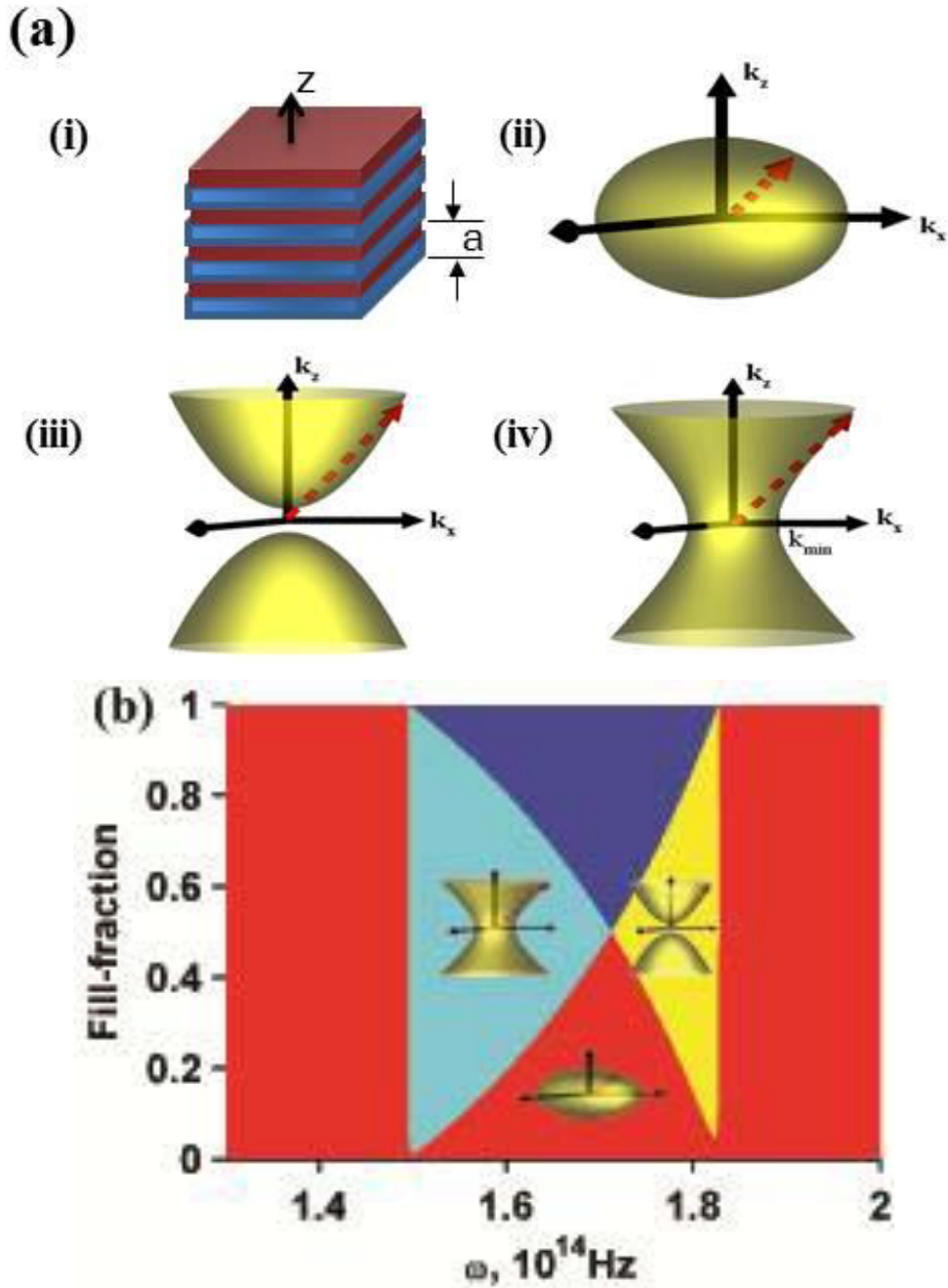

Figure 1: (a) (i) Multilayer hyperbolic metamaterial (HMM) with unit cell size 'a' consisting of alternating layers of silicon carbide and silicon dioxide (ii) Ellipsoidal isofrequency surface for effective anisotropic dielectric ($\varepsilon_{xx} = \varepsilon_{yy} > 0, \varepsilon_{zz} > 0$) (iii) type I HMM with only one negative component in the dielectric tensor ($\varepsilon_{xx} = \varepsilon_{yy} > 0, \varepsilon_{zz} < 0$) (iv) type II HMM with two negative components ($\varepsilon_{xx} = \varepsilon_{yy} < 0, \varepsilon_{zz} > 0$) The arrows denote the allowed wavevectors in the medium which can take large values in the HMM. (b) Optical phase diagram of SiC/SiO$_2$ metamaterial showing the different optical isofrequency surfaces achieved in different regions depending on the frequency of operation and fill fraction of metal. The dark blue area denotes an anisotropic effective metal where propagating waves are not allowed.

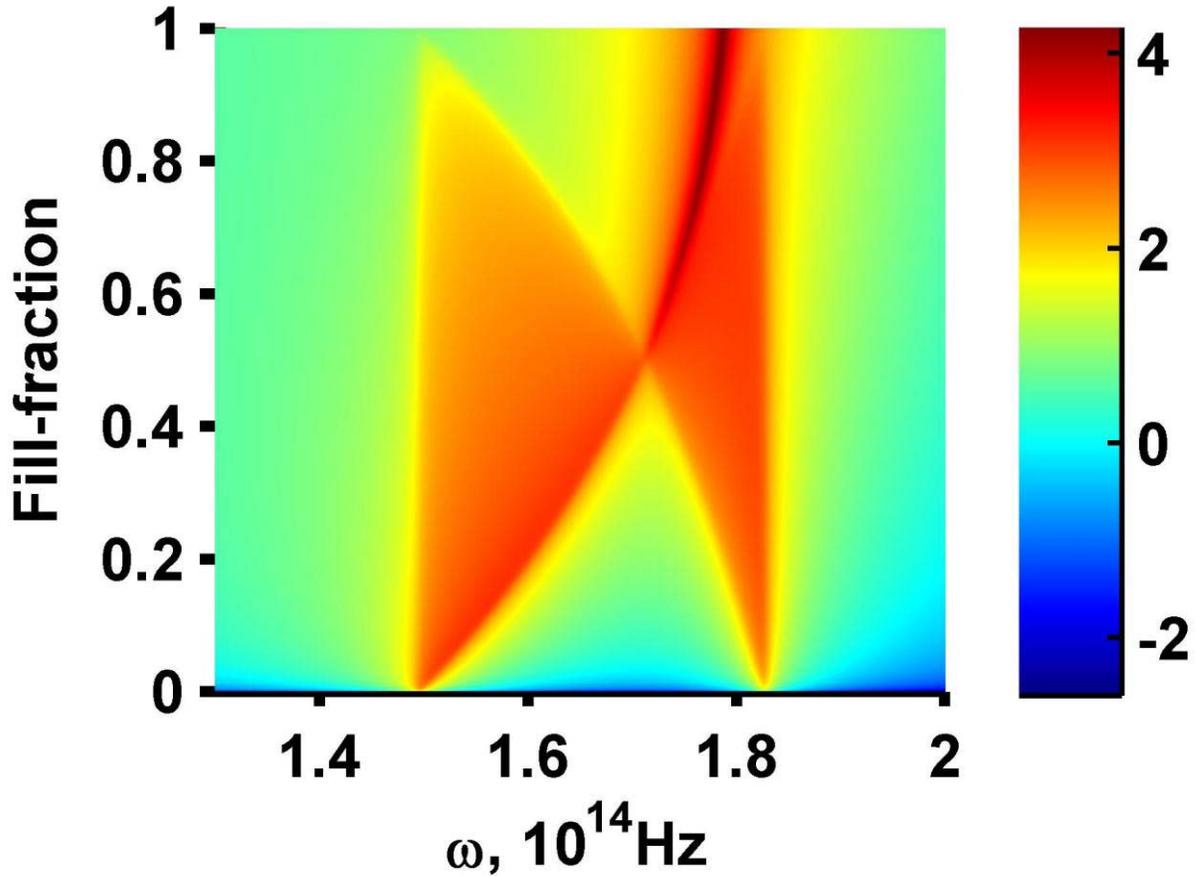

Figure 2: Super-planckian thermal emission in the near field of the HMM (normalized to the black body radiation into the upper half-space) calculated analytically. The characteristic dispersion of high-$k$ states and optical phases of the HMM are evident. Note that the thermal emission is enhanced in a broadband range due to the high-$k$ states present in the type I and type II HMM region but not in the effective dielectric region. The dark narrow band of high thermal emission occurs in the effective metal region due to a surface phonon polariton resonance.

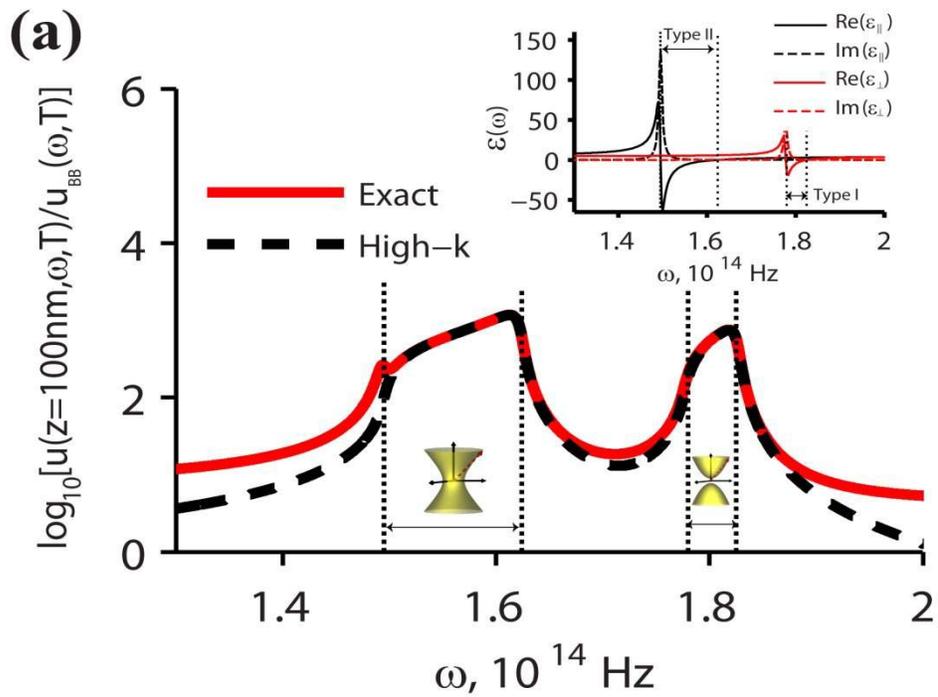

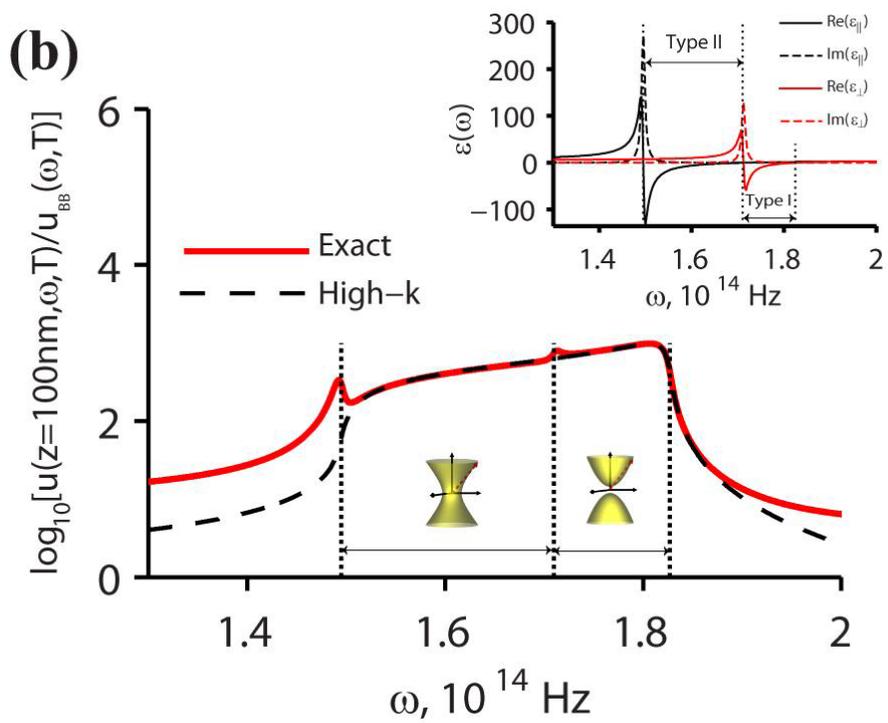

Figure 3: Comparison of the analytical result in the near field with an exact numerical solution for an effective medium HMM with metallic fill fractions (a) 0.25 and (b) 0.5. The thermal emission is normalized to the black body emission to the upper half-space with insets showing the effective medium dielectric parameters of the metamaterial. Note the excellent agreement between the analytical expression and full calculation. Furthermore, the broadband super-planckian thermal emission shows the expected topological transitions and closely follows the optical phases as expected. The two peaks in the exact calculation (b) correspond to the transverse optical phonon resonance ($\omega_{TO}$) and the singular meeting point of the optical phases ($\omega_s$).

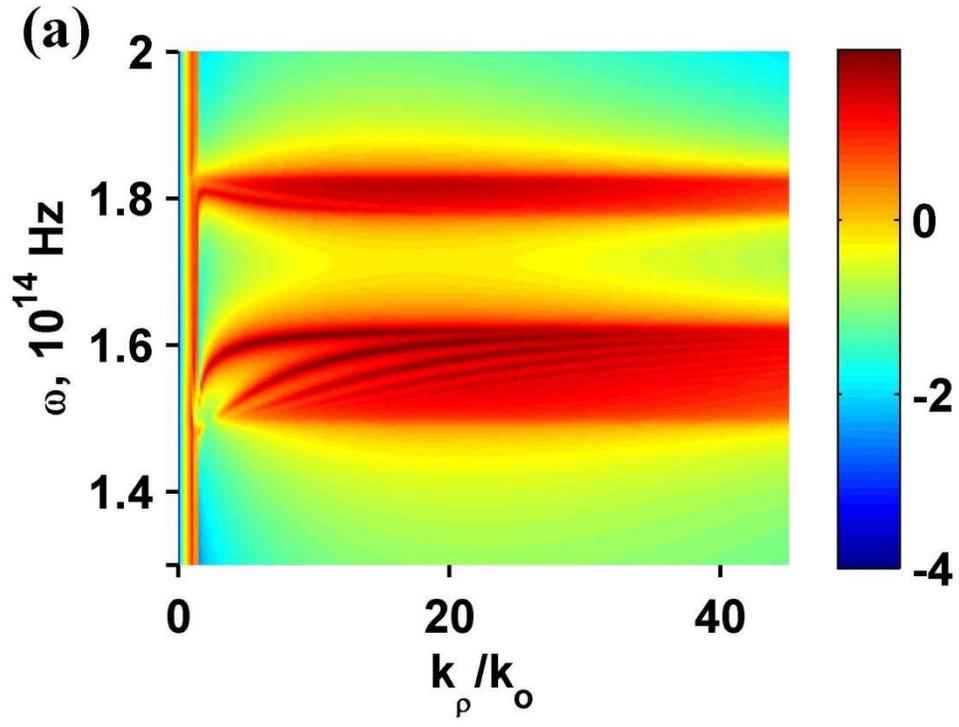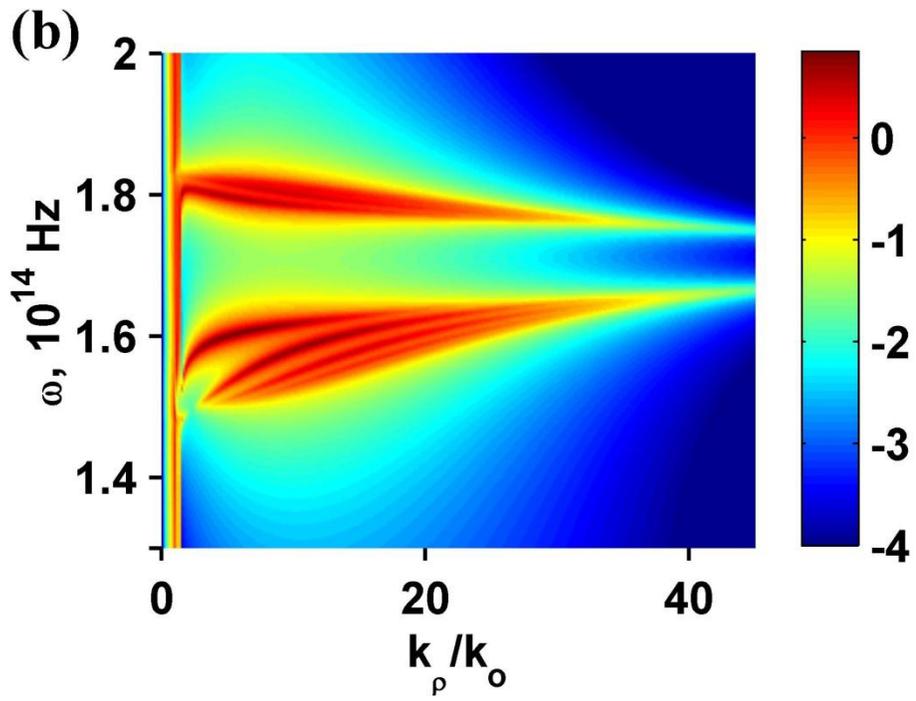

Figure 4: Comparison of the wavevector-resolved local density of states (WLDOS) which governs thermal emission (a) EMT prediction (b) Practical multilayer realization for f=0.25 and z=100nm. The structure consists of 10 layers of SiC/$SiO_2$ , 50nm/150 nm achieving a net thickness of 1 $\mu m$. The effect of the topological transitions as well as the presence of high-$k$ modes are clearly evident in the multilayer practical realization which takes into account all non-idealities due to dispersion, losses, finite unit cell size and finite sample size. The bright bands denote the enhanced LDOS due to high-$k$ modes in the HMM.

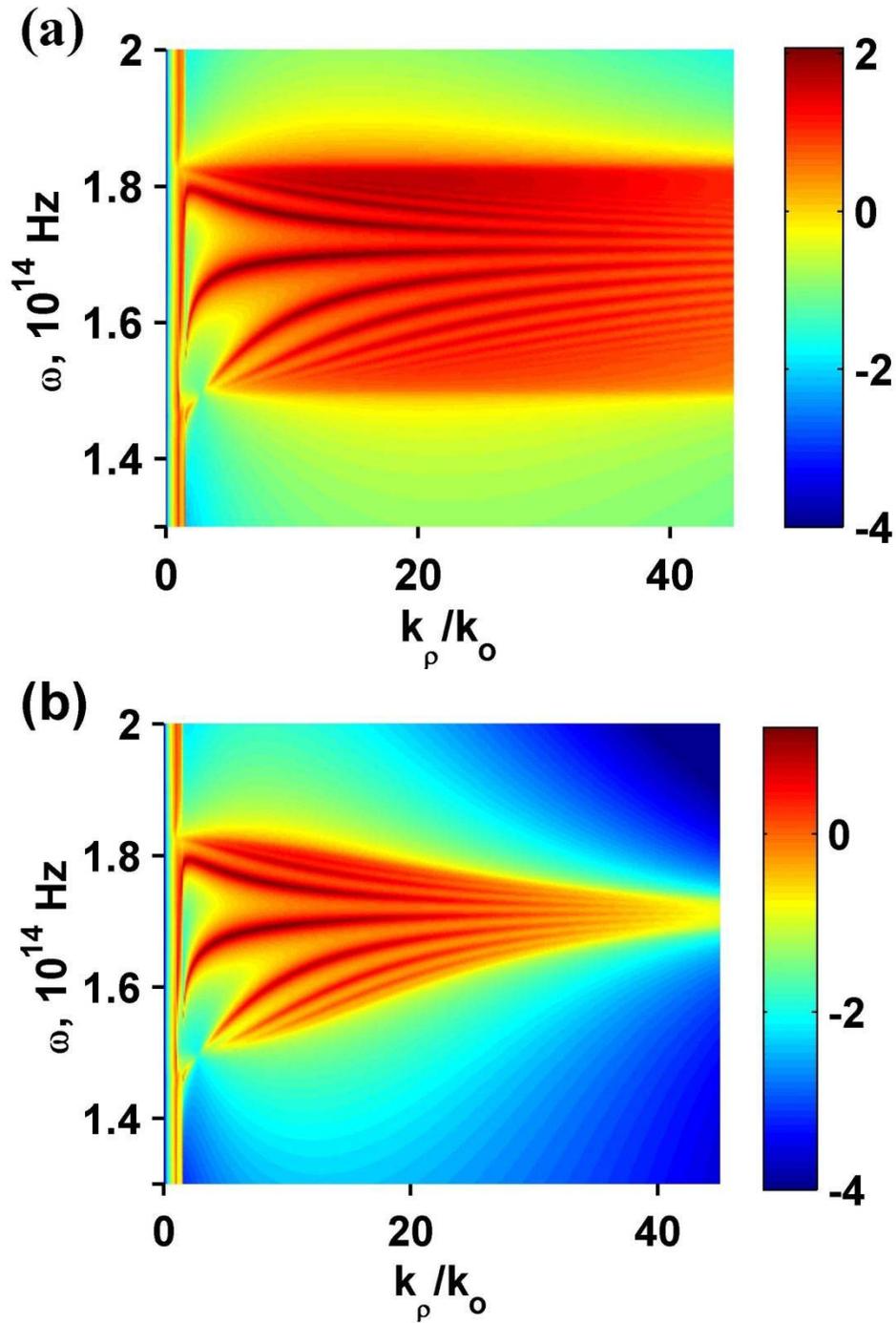

Figure 5: Broadband LDOS enhancement for f=0.5 (a) EMT prediction (b) practical multilayer structure with 10 layers of $SiC/SiO_2$, 100nm/100 nm achieving a net thickness of 1 $\mu m$. An upper cut off to the maximum wavevector exists in the multilayer realization but the net LDOS enhancement (and hence the super-planckian thermal emission) at a distance of z=100nm is over two orders of magnitude more than vacuum (see Fig. 3(b)).

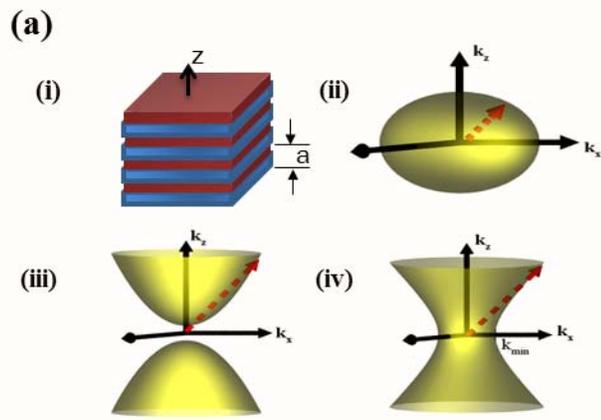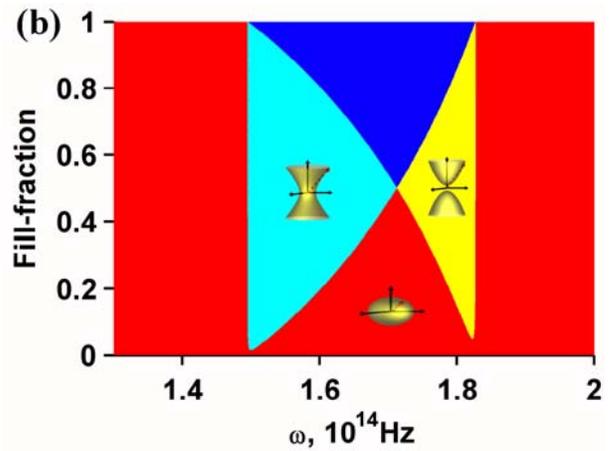

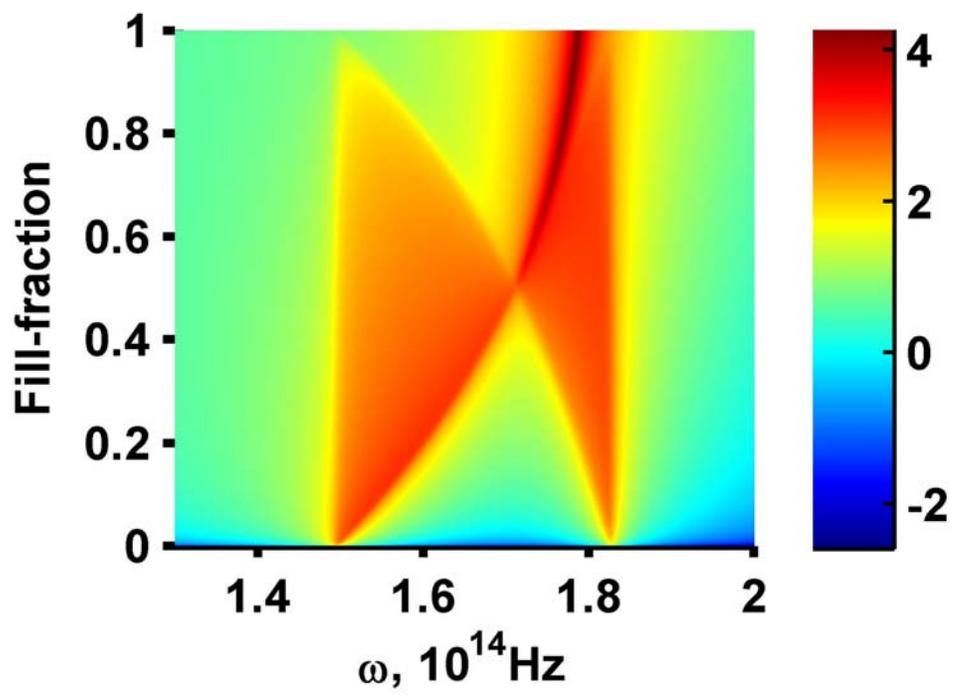

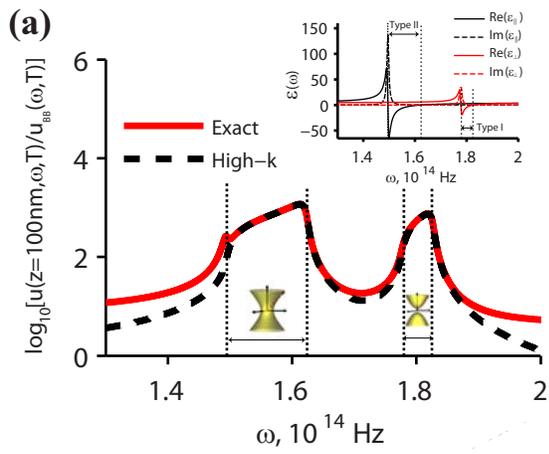 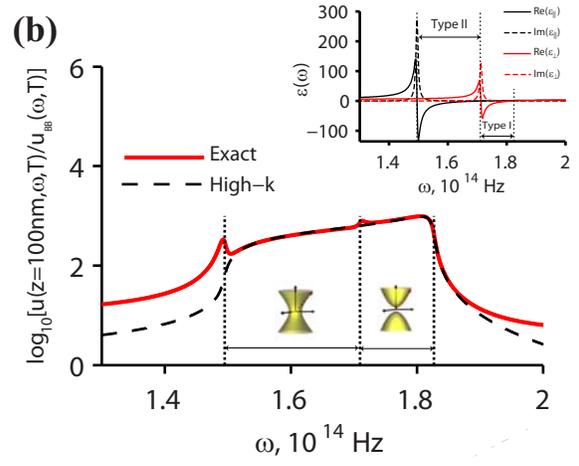

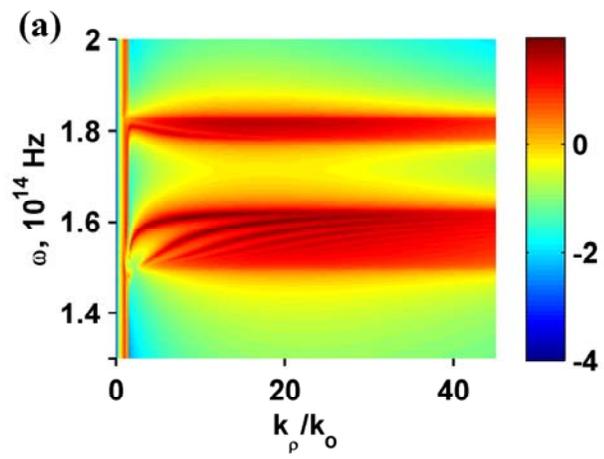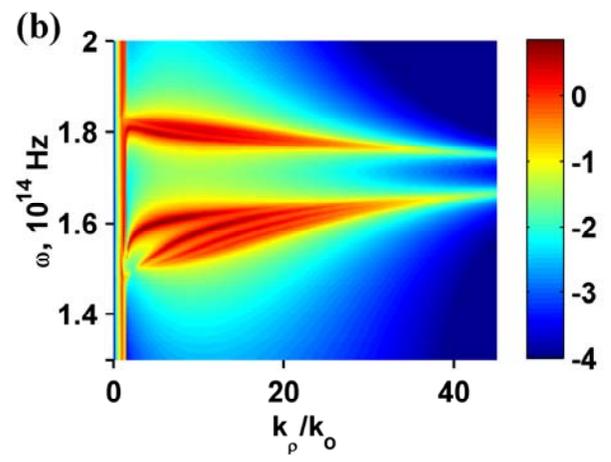

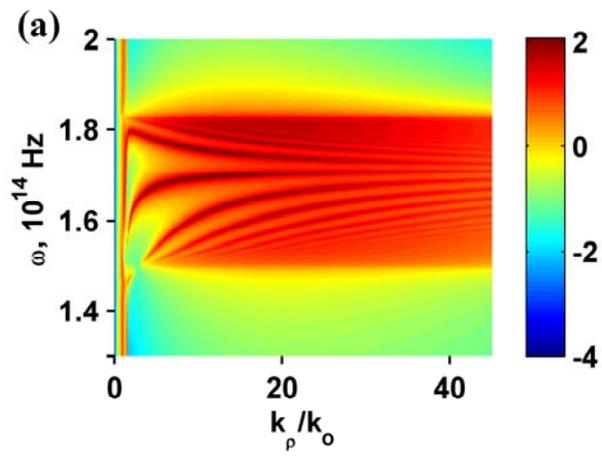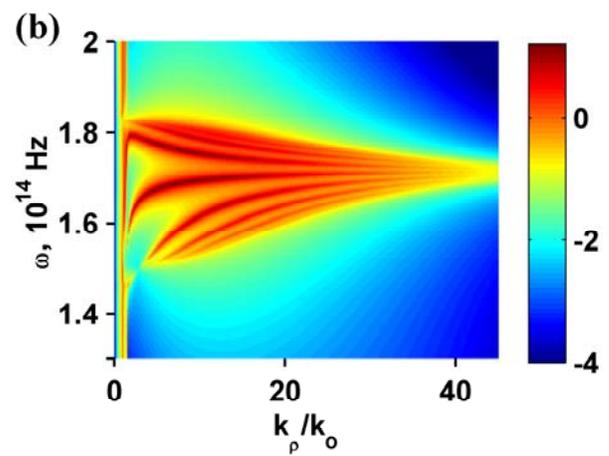